\documentclass[conference]{IEEEtran}
\IEEEoverridecommandlockouts
\usepackage{cite}
\usepackage{amsmath,amssymb,amsfonts}
\usepackage{algorithmic}
\usepackage{graphicx}
\usepackage{textcomp}
\usepackage{xcolor}
\usepackage{url} 
\usepackage{multirow}
\usepackage{graphicx}
\usepackage{multirow}
\usepackage[caption=false,font=footnotesize]{subfig}
\usepackage{hyperref}

\def\BibTeX{{\rm B\kern-.05em{\sc i\kern-.025em b}\kern-.08em
    T\kern-.1667em\lower.7ex\hbox{E}\kern-.125emX}}
\begin{document}

\newcommand{\old}[1]{\textcolor{green}{#1}}

\newcommand{\new}[1]{#1}

\newcommand{\todoadd}[1]{\textcolor{blue}{#1}}

\title{Evaluating Runtime Prediction and Conservative Walltime Policies in HPC Systems\\
}

\author{
\IEEEauthorblockN{
Anonymized
}
}
\author{
\IEEEauthorblockN{
Hector Romero-Ugalde
}
\IEEEauthorblockA{
\textit{R\&D Department, BULL} \\
Echirolles, France \\
0000-0002-7614-6474 \\
hector.romerougalde@bull.com
}
\and
\IEEEauthorblockN{
Youssef Faqir-Rhazoui
}
\IEEEauthorblockA{
\textit{R\&D Department, BULL} \\
Madrid, Spain \\
0000-0001-7833-1086 \\
youssef.el@bull.com
}
\and
\IEEEauthorblockN{
Marlon Funk
}
\IEEEauthorblockA{
\textit{R\&D Department, BULL} \\
Madrid, Spain \\
0009-0007-8183-2667 \\
marlon.funk@bull.com
}
\and
\IEEEauthorblockN{
Jesus Gorro\~nogoitia Cruz
}
\IEEEauthorblockA{
\textit{R\&D Department, BULL} \\
Madrid, Spain \\
jesus.gorronogoitia@bull.com
}
}

\maketitle

\IEEEpubid{%
  \makebox[\columnwidth][l]{%
    \parbox{\columnwidth}{%
      \raggedright
      \scriptsize
      Accepted for publication at IEEE/SBC SBAC-PAD 2026.\\
      \copyright\ 2026 IEEE. Personal use of this material is permitted.
    }%
  }%
  \hspace{\columnsep}%
  \makebox[\columnwidth]{}%
}

\begin{abstract}
Efficient use of large-scale supercomputing systems depends on accurate runtime estimation and walltime selection. To avoid premature termination, users often overestimate walltime, leading to poor resource utilization and reduced scheduling efficiency. \new{This paper presents a predictor-agnostic framework for safe walltime recommendation in production HPC environments. Rather than proposing a new runtime prediction model, the primary contribution is the SAFE decision layer, which converts predicted runtimes into conservative bounded walltime recommendations suitable for deployment. Runtime prediction is therefore treated as a modular component within the proposed framework.} We evaluate the approach using large-scale production job histories from the Fugaku and NREL Eagle supercomputers, comprising approximately 24M and 11M jobs, respectively. \new{Runtime prediction performance is first evaluated using representative predictors, including linear regression, neural networks, and a neural ensemble. The proposed SAFE framework is then evaluated using the NN ensemble predictor.}
Experimental results indicate that similarity-aware prediction substantially improves runtime estimation for workloads similar to previous executions. When combined with the proposed conservative walltime recommendation policies, these improvements enable walltime gap reductions of approximately 20 -- 26\% while maintaining job termination rates below 1\% across both systems.

\end{abstract}


\section{Introduction}
\label{sec:intro}
\IEEEpubidadjcol
High-performance computing (HPC) centres allocate resources through schedulers that require users to specify a walltime, i.e., the maximum runtime allowed for a job. When no value is provided, a system-defined default walltime is applied, which is typically not adapted to the characteristics of the submitted job. Walltime is a critical parameter: if it is set too low, jobs are terminated before completion; if it is set too high, cluster utilization degrades through inflated reservations, resulting in reduced backfilling opportunities and longer queue times. In many production environments, users adopt conservative habits and systematically over-request walltime to avoid failures, which creates persistent inefficiencies at scale \cite{zaborovsky2023improving}.

Accurate runtime prediction has therefore been studied extensively as a way to improve scheduling decisions and user guidance \cite{gaussier2015improving}. However, directly substituting a predicted runtime for walltime may not be acceptable in production due to prediction errors, workload drift, and heterogeneous user behaviors which can lead to under-allocation events that are operationally costly and weaken user trust \cite{li2021okcm}. A practical solution must therefore deliver safe recommendations, reducing systematic over-allocation while keeping the probability of premature termination extremely low, and must remain robust as workloads evolve over time.

\new{This paper studies a predictor-agnostic framework for safe walltime recommendation in production HPC systems. Runtime prediction is treated as a modular component that estimates job duration, whereas the primary contribution is the SAFE decision layer that converts these predictions into conservative bounded walltime recommendations suitable for operational deployment. The proposed framework integrates (i) runtime prediction trained on recent job histories, (ii) similarity-aware applicability constraints that distinguish between previously observed and unseen workloads, and (iii) conservative adjustment rules that bound reductions of the user-requested walltime.}

We evaluate the approach using complete job histories from two large-scale systems, Fugaku \cite{antici_2024_11467483} and Eagle \cite{OEDI_Dataset_5860}, which collectively comprise more than 35 million executed jobs. To mimic realistic deployment conditions, we adopt a rolling temporal evaluation protocol in which models are trained on recent history and evaluated on future workloads.

The contributions of this work are threefold. First, we introduce similarity-aware prediction regimes that distinguish between workloads previously observed during training and previously unseen workloads, enabling a more realistic analysis of prediction reliability. Second, we propose the SAFE walltime adjustment rule, which converts predicted runtimes into conservative walltime recommendations through bounded reductions of the user-requested walltime. Third, we introduce a similarity-aware extension of SAFE that applies additional safety margins to previously unseen workloads.

The remainder of the paper is organized as follows. Section~\ref{sec:soa} reviews related work on runtime prediction and walltime selection. Section~\ref{sec:data} describes the datasets, proposed methods, and experimental environment. Section~\ref{sec:results} presents experimental results on prediction accuracy and walltime efficiency–safety trade-offs. Section~\ref{sec:discussion} discusses limitations of this work. 
Section~\ref{sec:conclusion} concludes and outlines directions for future work.
\section{Context Overview}
\label{sec:soa}
Runtime prediction for batch-scheduled HPC systems has been studied for a long time as a way to improve scheduling efficiency, backfilling, and overall resource utilisation. Early work already identified systematic user overestimation of walltime and demonstrated that historical execution data could be exploited to estimate job duration more accurately \cite{zaborovsky2023improving}.

\subsection{HPCs and Available Datasets}
The first aspect to consider is the data we can access and its quality. Since many HPC systems are closed environments, it is often difficult to access real user data. However, Feitelson et al. (2014) in \cite{FEITELSON20142967} compile many of the datasets used in this field. Unfortunately, most of the datasets presented in that work suffer from either a) no longer being available or b) being outdated.

Menear et al. \cite{menear2023mastering} showed that many papers on job duration prediction suffer from data replication issues because the datasets used are not publicly available or the links to them are no longer accessible.

In \cite{antici_2024_11467483}, Antici et al. (2024) present a comprehensive dataset from the Fugaku HPC system (ranked 7th in the TOP500 list of November 2025). The dataset spans from March 2021 to April 2024, comprising 24 million jobs.

Researchers from the NREL Eagle supercomputer (ranked 284th in the TOP500 list of November 2025) have also published a large dataset in \cite{OEDI_Dataset_5860}. This dataset contains approximately 11 million jobs collected between November 2018 and February 2023.

\subsection{Early Workload Analysis and Similarity-Based Approaches}
One of the first works focused on analysing HPC usage behaviour through clustering and workload characterisation was Zhang et al. (2012) \cite{zhang2012hpc}. They grouped historical jobs using clustering approaches such as K-Means, hierarchical clustering, and neural‑networks, and then used cluster‑level statistics, rather than learned prediction models, to estimate the runtime of newly submitted jobs. Their results showed that incorporating similarity via clustering significantly improves runtime estimation in heterogeneous HPC workloads. Related studies explored machine learning‑based strategies to support scheduling decisions, such as classifying short-long jobs \cite{zrigui2021understanding}, or predicting queue waiting times instead of job duration \cite{vercellino2023machine}. \new{More recently, Cui et al. (2024) \cite{cui2024clustering} proposed a clustering-based runtime prediction framework for backfilling that groups jobs according to behavioral similarity before estimating runtime. Their work further demonstrates that exploiting workload similarity remains an effective strategy for improving runtime estimation in heterogeneous HPC environments.
}

Together, these studies establish a key insight that persists across the literature: job similarity matters, and ignoring workload heterogeneity can lead to unstable predictions.

\subsection{Runtime Regressors and Encoders}
Chen et al. (2020) \cite{chen2020runtime} conducted a comprehensive empirical evaluation of several regression and ensemble methods, including linear regression, random forests, XGBoost, and LightGBM. They reported an $R^2 = 0.86$ with the LightGBM model across the three datasets used. However, there are concerns regarding the methodology, since the datasets employed range from 1996 to 2006, and the authors do not specify how the data were split for training and testing. This detail is crucial for realistic evaluation scenarios \cite{menear2023mastering}.

Tanash et al. (2021) \cite{tanash2021ampro} proposed AMPRO-HPCC, a user-oriented machine-learning tool for predicting job runtime and memory usage from Slurm accounting logs. Evaluated on approximately 17.6 million jobs from the Beocat supercomputing cluster, their Random Forest-based runtime predictor achieved $R^2=0.74$. However, in this case, there are also doubts about the random split of training and testing data \cite{menear2023mastering}.

Even though the main objective of Francesco Antici et al. (2025) \cite{antici2025f} was not to develop the best runtime predictor, they achieved a reasonable performance of $MAE \approx 200$ minutes on the Fugaku dataset, which is one of the most comprehensive datasets available. The key contribution of Antici et al. was the processing of system logs using techniques from Natural Language Processing (NLP). They concatenate all relevant system features into a single long string and then encode it using the Sentence-BERT (SBERT) encoder. The results obtained, compared with those of simpler encoders, were better. Since the Fugaku dataset is one of the inputs to our work, we adopt this approach as it has demonstrated its effectiveness. 

\new{
This line of research has continued to evolve in recent years. Hai et al. (2026) \cite{hai2026fine} fine-tune Sentence-BERT specifically for HPC job outcome prediction from textual job metadata, further demonstrating the effectiveness of embedding-based representations for HPC workloads. Similarly, Liu et al. (2025) \cite{liu2025ora} propose ORA, an online retrieval-augmented language model that incorporates information retrieved from job scripts in addition to metadata for runtime prediction, illustrating the continued evolution of language-model-based representations for HPC workload prediction.
}

\subsection{Runtime Prediction for Scheduling and System Optimisation}

Prior work on runtime prediction for HPC scheduling has shown that model design must account not only for predictive accuracy but also for the asymmetric impact of prediction errors on scheduling behaviour. Gaussier et al. (2015) \cite{gaussier2015improving} demonstrated that classical regression approaches optimised for symmetric error metrics are ill-suited to backfilling environments, and proposed online polynomial regression models trained with custom asymmetric loss functions that penalise under- and over-prediction differently. Their approach relies on submission-time features, including user and job history and resource requests, and updates model parameters incrementally to adapt to workload evolution.

The work highlights the operational importance of error asymmetry in scheduler-integrated runtime prediction. While Gaussier et al. directly inject predicted runtimes into the scheduling mechanism, our work adopts a deployment-oriented perspective in which runtime prediction is decoupled from the final walltime decision. Rather than substituting predictions directly, we explicitly bind and constrain their influence through similarity-aware applicability and conservative adjustment rules, addressing error asymmetry and workload drift in a user-facing, deployable manner. These approaches consistently highlight an important asymmetry: underestimation errors are far more costly than overestimation, potentially leading to premature job termination or missed deadlines. Nevertheless, most scheduler-centric methods assume that predictions can be used directly or that failures are acceptable, limiting applicability in conservative production environments.

Wang et al. (2021) \cite{wang2021rlschert} introduce a deep-reinforcement-learning job scheduler that consumes dynamic remaining-runtime predictions to guide scheduling actions, including job selection and termination. Evaluated on real job logs from the USTC Supercomputing Center, the approach demonstrates improvements over static heuristics and prior learning-based schedulers. In this work, runtime predictions are injected directly into the scheduling policy. The study reinforces our design choice to treat runtime prediction as an intermediate signal, explicitly decoupled from the final bounded walltime recommendation required for safe deployment.

\new{
Recent work has also incorporated uncertainty estimation into runtime prediction. For example, uncertainty-aware approaches combine multi-quantile regression with adaptive safety margins to estimate conservative runtime bounds while remaining compatible with existing schedulers \cite{choi2026uarp}. While these studies focus on improving runtime prediction accuracy or quantifying prediction uncertainty, our work addresses the complementary problem of converting runtime predictions into safe, deployable walltime recommendations through a predictor-agnostic decision framework.
}

\subsection{Temporal Validation and Concept Drift}

Menear et al. (2023) \cite{menear2023mastering} present a methodological framework for HPC runtime prediction that systematises dataset preparation, feature engineering, temporal validation, and error analysis, arguing that many published gains stem from evaluation leakage and non‑representative splits rather than truly generalizable models. They emphasise rolling, time‑aware validation, strong baselines, and transparent reporting (e.g., MAE and R2 with temporal holds), offering a reproducible rubric to compare models across workloads and institutions. This methodology was also adopted by other relevant works in the resource prediction subject \cite{gallo2025mitigating} 

This perspective aligns with our deployment‑oriented design: we adopt rolling-window training and next‑day evaluation to study workload drift, and pair accuracy metrics with safety‑aware walltime outcomes, thereby extending their methodological guidance from prediction rigour to conservative decision rules for user‑facing walltime recommendations.
\section{Methods, Data, and Procedures}
\label{sec:data}

\subsection{Data Description}\label{sec:datadata}
We conduct our study using production accounting logs collected from the Fugaku \cite{antici_2024_11467483} and NREL Eagle \cite{OEDI_Dataset_5860} supercomputers. Table \ref{tab:datasets} depicts the overall metrics collected from both datasets. Altogether, the datasets account for more than 35 million jobs. While Fugaku has almost 3.5K users, Eagle serves nearly 1K different users.
Note that, on average, each user accounts for approximately 
$7K$ jobs in Fugaku and $11K$ jobs in Eagle. This high level of workload repetition motivates several aspects of this study, including the similarity-aware prediction regimes and the analysis of previously observed versus unseen workloads.

Although both datasets use different workload managers (PBS vs. SLURM), most of the \textit{pre-run}\footnote{\textit{pre-run} metrics refer to those available when a job is submitted but before it is launched.} metrics are the same. These include the user name, job ID, number of requested nodes, requested memory, and maximum requested walltime, among others. Therefore, despite the different data sources, we created an intermediate interface to support both datasets. Regarding \textit{post-run} metrics, we only used the job duration to evaluate our models, and the exit status (or exit code) to filter the jobs.

Only successfully completed jobs were retained. Runtime values were converted to minutes and denoted as $T_\text{obs}$ (Time observed). To prevent temporal leakage and preserve realistic deployment conditions, all experiments use strictly time-ordered splits.

\begin{table}[!ht]
\centering
\resizebox{\columnwidth}{!}{%
\begin{tabular}{|lcccccc|}
\hline
                & \textbf{\#jobs} & \textbf{\#users} & \textbf{\#features}                                               & \textbf{time window}                                           & \textbf{\begin{tabular}[c]{@{}c@{}}workload\\ system\end{tabular}} & \multicolumn{1}{l|}{\textbf{\begin{tabular}[c]{@{}l@{}}total\\ size\end{tabular}}} \\ \hline
\textbf{Fugaku} & 24M             & 3,457            & \begin{tabular}[c]{@{}c@{}}pre-run: 9\\ post-run: 36\end{tabular} & \begin{tabular}[c]{@{}c@{}}Mar 2021 -\\ Apr 2024\end{tabular} & PBS                                                                & 28GB                                                                              \\ \hline
\textbf{Eagle}  & 11M             & 936              & \begin{tabular}[c]{@{}c@{}}pre-run: 13\\ post-run: 5\end{tabular} & \begin{tabular}[c]{@{}c@{}}Nov 2018 -\\ Feb 2023\end{tabular} & SLURM                                                              & 240MB                                                                             \\ \hline
\end{tabular}%
}
\caption{General statistics for the Fugaku and Eagle datasets.}
\label{tab:datasets}
\end{table}

\subsection{Methods}
\subsubsection{Temporal Rolling Evaluation}
We adopt a sliding-window temporal validation strategy to mimic online deployment. For each split: (1) Training window: 29 consecutive days, (2) Validation window: 1 day, and (3) Test window: 1 subsequent day.

In other words, the runtime of each job within the testing window (one day) is predicted, and its walltime is adjusted based on the predicted runtime obtained from the previous 30 days of job history and the user-requested walltime.

This rolling evaluation protocol simulates real-time prediction of unseen future workloads and avoids retrospective bias.

\subsubsection{Feature Representation}
Job metadata is represented using dense vector embeddings (emb), which encode semantic information derived from job names, user metadata, and user-requested resources (e.g., memory, nodes, GPUs, etc.).

Following the approach in \cite{antici_2024_11467483}, we concatenate all job features into a comma-separated string such as ``jobname,jobenv,...''. This string is then encoded using the SBERT model. The output is a 384-dimensional dense vector represented as 64-bit floating-point values, which we use as input to the models.

\subsubsection{Predictive Models}
In \textit{Neural Networks}(NNs), for each split, we train feed-forward multilayer perceptrons (MLPs) to predict job runtime. Input features correspond to 384-dimensional SBERT embeddings derived from job metadata.

The evaluated MLP architectures consist of fully connected networks with three hidden layers. Hidden dimensions vary across candidate configurations, including structures such as $(80,160,80)$, $(100,200,100)$, $(120,240,120)$, $(150,200,150)$, and $(256,128,64)$. Activation functions include combinations of \texttt{tanh} and \texttt{ReLU}. The output layer contains a single neuron for runtime regression.

Training is performed using PyTorch with mini-batch gradient descent. For each temporal split, multiple candidate configurations are trained independently using different combinations of architectures, activation functions, batch sizes, training epochs, and loss functions. Candidate models are trained for 20 to 50 epochs using batch sizes of 64, 128 and 1024 and a learning rate fixed at $10^{-3}$. Different candidate configurations are trained using either Mean Squared Error (MSE) loss or Huber loss. Model selection is performed according to validation RMSE. After selecting the best-performing configurations, the models are retrained on the combined training and validation sets before final test evaluation.

\textit{Ensemble NNs: } To improve robustness and reduce variance, we construct an ensemble from the top $M=5$ validation-performing models. Given predictions $\hat{T}_m$ from each model:
\begin{equation}\label{pred}
\hat{T}_{ens}=mean(\hat{T}_1,\cdots, \hat{T}_M)
\end{equation}

In practice, ensemble predictions are combined using inverse-validation-RMSE weighting, giving larger influence to models with lower validation error. We report results for both the single best model and the ensemble.

\textit{Linear Regression Baseline:} As a lightweight baseline, we train an ordinary least squares linear regression model on the same feature space. This provides a parametric reference to quantify the benefit of nonlinear modelling and ensembling.

\new{The objective of this comparison is not to perform an exhaustive benchmark of runtime prediction algorithms, but rather to evaluate the proposed SAFE framework using representative predictors of increasing complexity. Because SAFE is predictor independent, alternative regression models can be incorporated without modifying the proposed walltime recommendation policy.}

\subsubsection{Similarity-Aware Prediction Regimes}\label{similaritysubsection}

Production HPC workloads frequently contain repeated executions of the same or highly similar job scripts. As a result, job runtime prediction accuracy may differ depending on whether the job is similar to workloads previously observed during training. To capture this effect, we introduce three prediction regimes:

\begin{itemize}

\item \textbf{ALL}: predictions evaluated over the complete test set, including both previously observed and previously unseen workloads.
\item \textbf{IN}: predictions restricted to jobs whose feature signatures were previously observed in the training data.
\item \textbf{OUT}: jobs whose feature signatures were not observed during training.

\end{itemize}

By definition, the ALL regime therefore corresponds to the union of IN and OUT jobs.

A job is classified as \textbf{IN} if its feature signature was previously observed in the training set; otherwise it is considered \textbf{OUT}. 

We expect runtime predictions to be more accurate for \textbf{IN} jobs, since the model has already observed similar workload patterns during training.

In our deployment scenario, models are retrained periodically using a rolling temporal window of 30 days. When a new job arrives, its feature signature is compared against the signatures present in the training set. If the job has been previously observed, the prediction is considered \emph{confident} and falls into the \textbf{IN} regime. Otherwise, the job is classified as \textbf{OUT}, indicating that the workload has not been seen during training and therefore the prediction confidence is lower.

Evaluating prediction performance separately for these regimes provides insight into how runtime prediction behaves under realistic production conditions, where systems continuously encounter both repeated workloads and novel applications.

\subsubsection{Walltime Gap Reduction}

While runtime prediction can provide useful estimations of job duration, directly using predicted runtimes in scheduling decisions may lead to unsafe underestimation of execution time. HPC schedulers therefore require conservative runtime estimates to avoid premature job termination.

To address this issue, we evaluate a conservative heuristic adjustment rule that converts the predicted runtime $\hat{T}$ into a walltime recommendation $\hat{T}_{adj}$:

\begin{equation}\label{rulewall}
\hat{T}_{adj} =
\begin{cases}
\max(\beta \cdot elpl, \hat{T}) & \text{if } \frac{\hat{T}}{elpl} < \alpha \\
elpl & \text{otherwise}
\end{cases}
\end{equation}

where $elpl$ denotes the user-requested walltime. The parameter $\alpha$ defines the threshold below which the predicted runtime is considered sufficiently smaller than the requested walltime to justify a reduction. The parameter $\beta$ determines the fraction of the original walltime retained when a reduction is applied, while the $\max(\cdot)$ operator ensures that the recommended walltime never falls below the predicted runtime.

In our experiments we set $\alpha = 0.1$ and $\beta = 0.5$, meaning that reductions are triggered only when the predicted runtime is less than 10\% of the requested walltime, and the adjusted walltime retains half of the original request. 
These parameters were intentionally selected conservatively in order to prioritize low job termination rates over aggressive walltime reduction. The objective of this work is not to optimize these hyperparameters, but rather to evaluate whether simple bounded adjustment policies can provide stable operational trade-offs across large-scale production workloads.

\subsubsection{Similarity-aware safety adjustment.}

The previous rule treats all predictions equally. However, as discussed in Section~\ref{similaritysubsection}, prediction reliability differs depending on whether a job has been previously observed during training (\textbf{IN}) or corresponds to a previously unseen workload (\textbf{OUT}). Predictions for OUT jobs are generally less reliable because the model must extrapolate beyond previously observed workload patterns.

To account for this difference in prediction confidence, we introduce a similarity-aware adjustment rule that applies an additional safety margin to OUT jobs. Specifically, predicted runtimes for OUT jobs are first inflated by a multiplicative safety factor $\gamma$, as defined in (\ref{paperule}), before applying the SAFE adjustment rule (\ref{rulewall}).

\begin{equation} \label{paperule}
\hat{T}^{*} =
\begin{cases}
\hat{T} & \text{if job is IN} \\
\gamma \cdot \hat{T} & \text{if job is OUT}
\end{cases}
\end{equation}

where $\gamma > 1$ increases the predicted runtime for previously unseen workloads. \new{Following \cite{vallabhajosyula2022towards}, we adopt $\gamma=1.5$ as a conservative operating point for previously unseen workloads. The objective of this work is not to optimize this parameter, but rather to evaluate whether a simple similarity-aware safety margin improves the robustness of walltime recommendations for previously unseen workloads. Automatic calibration of $\gamma$ is left for future work.}

This hybrid rule combines similarity-aware prediction regimes with the SAFE walltime adjustment mechanism. In practice, the scheduler applies the standard SAFE rule for repeated workloads (IN), while applying a more conservative correction for previously unseen jobs (OUT).

\subsubsection{Evaluation Metrics}
The prediction accuracy is evaluated on each test day using the Root Mean Square Error (RMSE), the Mean Absolute Error (MAE) and the coefficient of determination ($R^2$). The metrics are computed independently for the best NN model, the NN ensemble, and a linear model regression baseline.  

In addition to prediction accuracy, we evaluate the impact of runtime prediction on walltime estimation. The walltime gap is defined as
\begin{equation}
\Delta_{\text{wall}} = elpl - T_{\text{obs}}
\end{equation}

where $elpl$ is the user-requested walltime and $T_{\text{obs}}$ is the observed runtime. Using the adjusted prediction $\hat{T}_{adj}$ introduced in \ref{rulewall}, we compute the predicted gap

\begin{equation}
\Delta_{\text{pred}} = \hat{T}_{adj} - T_{\text{obs}}.
\end{equation}

The potential benefit of runtime prediction is measured as the reduction in overestimation:

\begin{equation}
\text{Saved time} =
\sum_{i=1}^{n} \begin{cases}
\Delta_{\text{wall}_i} - \Delta_{\text{pred}_i} & i_{\text{pred}} \notin \mathrm{C}\\
\Delta_{\text{wall}_i} & i_{\text{pred}} \in \mathrm{C}
\end{cases}
\end{equation}

where $\mathrm{C}$ is the set of jobs cut by the prediction at the testing time. Since model predictions can lead to early job termination, we consider only cases where the job was not terminated. Otherwise, we accumulate the base walltime, assuming the user will re-run the same job.

All temporal splits are evaluated independently and aggregated to obtain global statistics during the evaluation period.

\subsection{Experimental Environment}
Due to the overhead of computing SBERT embeddings on the fly, we pre-computed them for both datasets using an AMD Instinct MI210 with ROCm 6.3.3 drivers. For the SBERT implementation, we used the \textit{all-MiniLM-L6-v2}\footnote{\url{https://huggingface.co/sentence-transformers/all-MiniLM-L6-v2}}
 model from Hugging Face, as well as the AMD version of PyTorch\footnote{\url{https://pytorch.org/get-started/locally/}}

Regarding the offline evaluation, we conducted the experiments on a cluster of six nodes equipped with Intel Xeon 6426Y processors. We split the workload into batches to utilize each compute node and accelerate the model training.

The source code is publicly available on GitHub.\footnote{
\href{https://github.com/hromerougalde/safe-walltime-recommendation/tree/sbacpad2026}
{SAFE Walltime Recommendation -- SBAC-PAD 2026}
}
\section{Experimental Results}
\label{sec:results}

\subsection{Prediction performance}

Table~\ref{tab:prediction_metrics_combined} reports the results for the different models on both systems. The objective of this comparison is to evaluate whether prediction performance differs between the IN subset and the full workload (ALL). Reporting the metrics on both scopes allows a direct comparison of model accuracy for previously observed workloads (IN) and for the complete test workload (ALL).

\begin{table}[!ht]
\centering
\begin{tabular}{lllrrr}
\hline
\textbf{System}         & \textbf{Scope}       & \textbf{Model}        & \textbf{RMSE}   & \textbf{MAE}   & \textbf{R2}   \\ \hline
\multirow{6}{*}{Fugaku} & \multirow{3}{*}{ALL} & LINEAR                & 317.61          & 167.44         & 0.32          \\
                        &                      & NN\_BEST              & 257.61          & 88.89          & 0.55          \\
                        &                      & \textbf{NN\_ENSEMBLE} & \textbf{250.65} & \textbf{87.64} & \textbf{0.57} \\ \cline{2-6} 
                        & \multirow{3}{*}{IN}  & LINEAR                & 234.67          & 107.94         & 0.35          \\
                        &                      & NN\_BEST              & 170.16          & 49.0           & 0.66          \\
                        &                      & \textbf{NN\_ENSEMBLE} & \textbf{168.53} & \textbf{48.70} & \textbf{0.67} \\ \hline
\multirow{6}{*}{Eagle}  & \multirow{3}{*}{ALL} & LINEAR                & 374.11          & 167.30         & 0.11          \\
                        &                      & NN\_BEST              & 349.28          & 99.16          & 0.23          \\
                        &                      & \textbf{NN\_ENSEMBLE} & \textbf{343.07} & \textbf{98.76} & \textbf{0.26} \\ \cline{2-6} 
                        & \multirow{3}{*}{IN}  & LINEAR                & 190.98          & 76.13          & 0.17          \\
                        &                      & NN\_BEST              & 146.71          & 42.87          & 0.51          \\
                        &                      & \textbf{NN\_ENSEMBLE} & \textbf{141.33} & \textbf{41.84} & \textbf{0.55} \\ \hline
\end{tabular}
\caption{Prediction accuracy of the evaluated models on the Fugaku and Eagle HPC systems.}
\label{tab:prediction_metrics_combined}
\end{table}

Across both systems, prediction performance is consistently better on the IN subset than on the full dataset (ALL). RMSE and MAE decrease substantially, while $R^2$ increases. On Fugaku, the MAE of the NN ensemble decreases from 87.6 minutes on ALL to 48.7 minutes on IN, with $R^2$ increasing from 0.57 to 0.67. A similar trend is observed on Eagle, where the MAE decreases from 98.8 minutes to 41.8 minutes and $R^2$ increases from 0.26 to 0.55. Linear regression follows the same pattern but with consistently higher errors.

\new{These results indicate that workload familiarity is a major determinant of runtime prediction accuracy in production HPC systems. The substantial improvement observed for repeated workloads (IN) motivates the similarity-aware SAFE policy introduced later, where prediction confidence directly influences the conservativeness of walltime recommendations.}

Across both systems, NN models outperform the linear baseline. On Fugaku, the NN ensemble reduces MAE from 167.4 minutes (linear) to 87.6 minutes, while on Eagle, it reduces MAE from 167.3 minutes to 98.8 minutes.

\begin{figure*}[!t]
\centering

\subfloat[NN vs Linear (Fugaku)]{
    \includegraphics[width=0.48\linewidth]{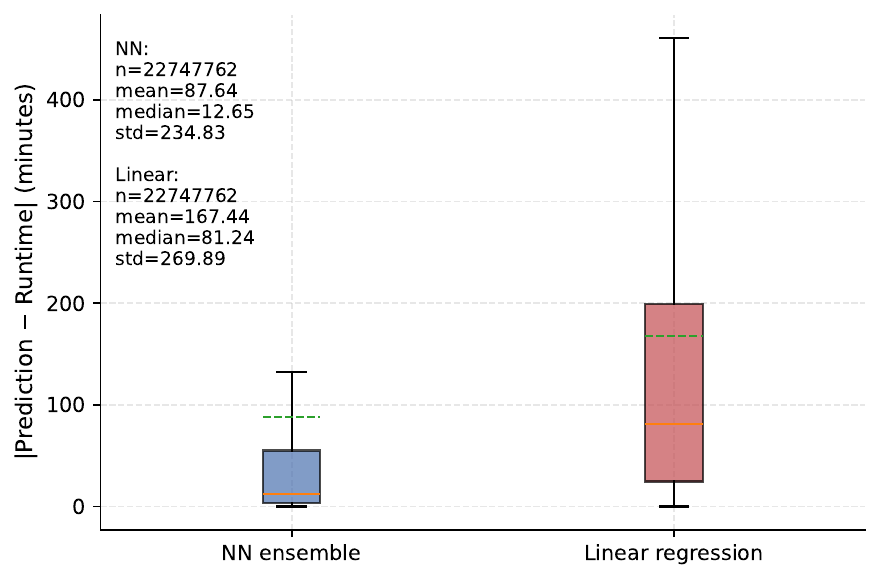}
    \label{fig:nn_vs_lr_fugaku}
}
\hfill
\subfloat[NN vs Linear (Eagle)]{
    \includegraphics[width=0.48\linewidth]{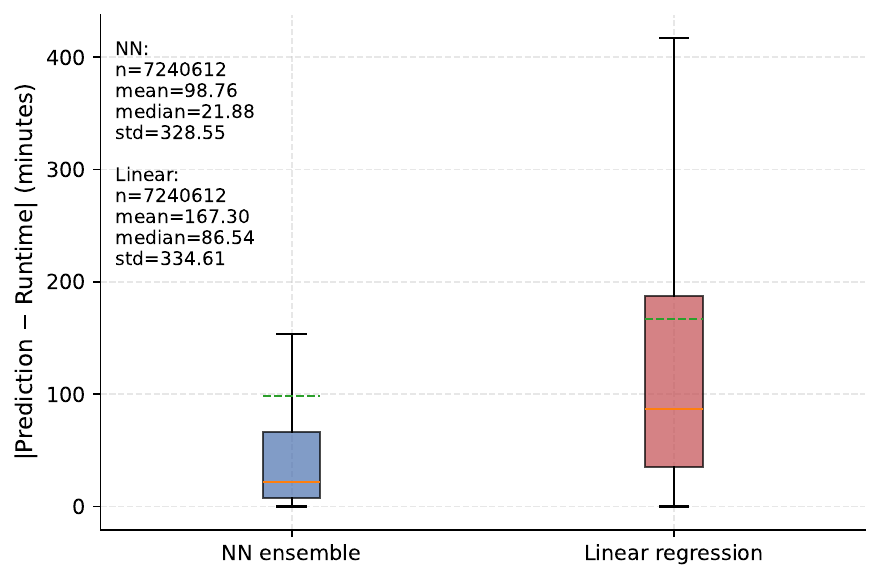}
    \label{fig:nn_vs_lr_eagle}
}

\vspace{0.15cm}

\subfloat[ALL vs IN prediction errors (Fugaku)]{
    \includegraphics[width=0.48\linewidth]{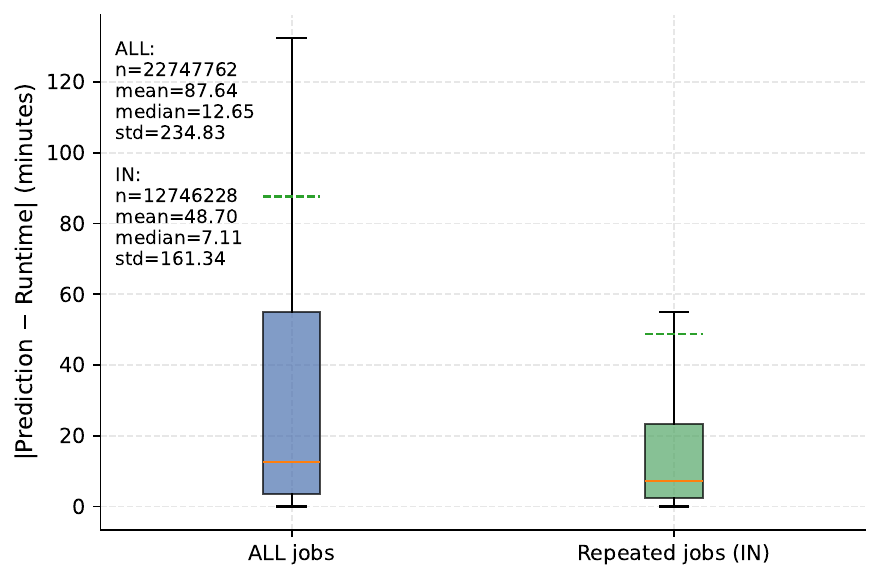}
    \label{fig:all_vs_in_fugaku}
}
\hfill
\subfloat[ALL vs IN prediction errors (Eagle)]{
    \includegraphics[width=0.48\linewidth]{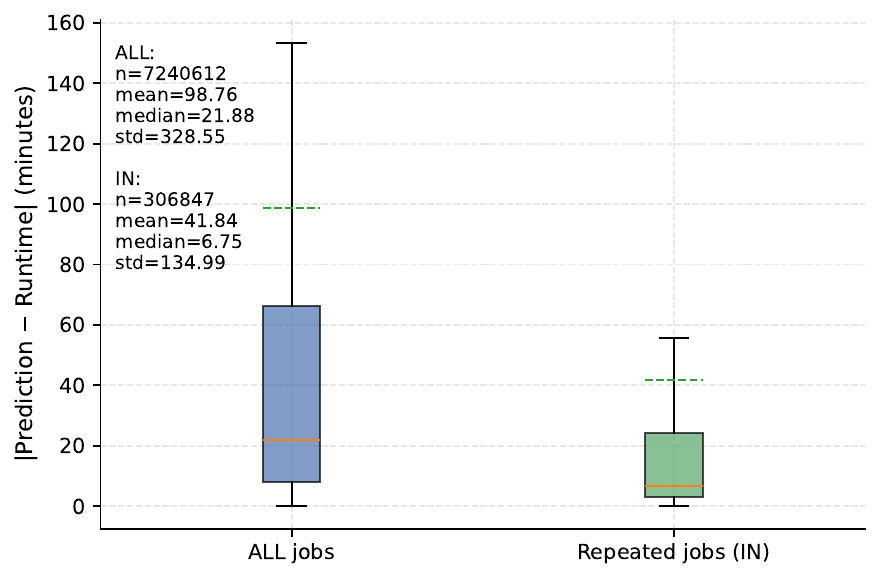}
    \label{fig:all_vs_in_eagle}
}

\caption{
Distribution of absolute prediction errors across models and prediction regimes.
(a,b) Comparison between NN ensemble and linear regression models.
(c,d) Comparison between ALL jobs and repeated workloads (IN).
}
\label{fig:prediction_error_comparison}

\end{figure*}

Figures~\ref{fig:prediction_error_comparison}a and~\ref{fig:prediction_error_comparison}b compare the distribution of absolute prediction errors for the NN ensemble and linear regression models on the Fugaku and Eagle systems. In both datasets, the NN ensemble exhibits substantially lower median and mean errors than the linear baseline, indicating improved prediction accuracy.

Figures~\ref{fig:prediction_error_comparison}c and~\ref{fig:prediction_error_comparison}d compare prediction errors between the ALL and IN regimes for the NN ensemble model. Consistent with the metrics reported in Table~\ref{tab:prediction_metrics_combined}, restricting evaluation to repeated workloads (IN) substantially reduces prediction errors relative to the full workload (ALL).

\subsection{Contributions 2 and 3: Impact of allocation efficiency}

Table~\ref{tab:allocation_results_combined} reports the effect of different allocation
policies on walltime gap reduction and job termination rates for the Fugaku and Eagle workloads. The table compares five approaches for reducing walltime overestimation: (i) direct allocation using predicted runtimes, (ii) a baseline strategy that scales the user-requested walltime without using runtime predictions, (iii) a safety-factor approach where predicted runtimes are multiplied by a constant factor to reduce the risk of job failures, following prior work on machine-learning based runtime prediction for HPC scheduling \cite{vallabhajosyula2022towards}, (iv) the proposed SAFE walltime adjustment rule (Contribution~2), and (v) the similarity-aware SAFE rule (Contribution~3).

\begin{table*}[!t]
\centering
\begin{tabular}{llllrr}
\hline
\textbf{System}         & \textbf{Model}                          & \textbf{Policy}                             & \textbf{Parameters}                     & \textbf{Gap Red. (\%)} & \textbf{Job Cuts (\%)} \\ \hline
\multicolumn{6}{c}{\textbf{Baselines: Direct allocation}}                                                                                                                                                   \\ \hline
\multirow{3}{*}{Fugaku} & Linear                                  & \multirow{6}{*}{Direct}                     & \multicolumn{1}{c}{\multirow{6}{*}{--}} & 54.36                  & 42.54                  \\
                        & NN best                                 &                                             & \multicolumn{1}{c}{}                    & 53.15                  & 44.54                  \\
                        & \textbf{NN ensemble}                    &                                             & \multicolumn{1}{c}{}                    & \textbf{55.55}         & \textbf{41.13}         \\ \cline{1-2} \cline{5-6} 
\multirow{3}{*}{Eagle}  & Linear                                  &                                             & \multicolumn{1}{c}{}                    & 54.92                  & 43.97                  \\
                        & NN best                                 &                                             & \multicolumn{1}{c}{}                    & 58.02                  & 37.06                  \\
                        & \textbf{NN ensemble}                    &                                             & \multicolumn{1}{c}{}                    & \textbf{61.82}         & \textbf{31.70}         \\ \hline
\multicolumn{6}{c}{\textbf{Baseline: Walltime scaling}}                                                                                                                                                     \\ \hline
\multirow{2}{*}{Fugaku} & \multicolumn{1}{c}{\multirow{4}{*}{--}} & \multirow{4}{*}{Walltime scaling}           & $0.80 \times elpl$                      & 26.10                  & 8.08                   \\
                        & \multicolumn{1}{c}{}                    &                                             & \textbf{$0.90 \times elpl$}             & \textbf{13.68}         & \textbf{2.96}          \\ \cline{1-1} \cline{4-6} 
\multirow{2}{*}{Eagle}  & \multicolumn{1}{c}{}                    &                                             & $0.80 \times elpl$                      & 22.75                  & 2.30                   \\
                        & \multicolumn{1}{c}{}                    &                                             & \textbf{$0.90 \times elpl$}             & \textbf{11.51}         & \textbf{1.18}          \\ \hline
\multicolumn{6}{c}{\textbf{Baseline: Prediction inflation\cite{vallabhajosyula2022towards}}}                                                                                              \\ \hline
Fugaku                  & \multirow{2}{*}{NN ensemble}            & \multirow{2}{*}{$\kappa \times$ prediction} & \multirow{2}{*}{$\kappa=1.5$}           & 63.89                  & 20.08                  \\ \cline{1-1} \cline{5-6} 
Eagle                   &                                         &                                             &                                         & 68.30                  & 20.92                  \\ \hline
\multicolumn{6}{c}{\textbf{Contribution 2: SAFE walltime adjustment}}                                                                                                                                       \\ \hline
Fugaku                  & \multirow{2}{*}{NN ensemble}            & \multirow{2}{*}{SAFE}                       & \multirow{2}{*}{$\alpha=0.1,\beta=0.5$} & 19.19                  & 1.10                   \\ \cline{1-1} \cline{5-6} 
Eagle                   &                                         &                                             &                                         & \textbf{31.63}         & \textbf{1.03}          \\ \hline
\multicolumn{6}{c}{\textbf{Contribution 3: Similarity-aware SAFE}}                                                                                                                                          \\ \hline
Fugaku                  & \multirow{2}{*}{NN ensemble}            & \multirow{2}{*}{Similarity-aware SAFE}      & \multirow{2}{*}{$\alpha=0.1,\beta=0.5$} & 19.71                  & 0.91                   \\ \cline{1-1} \cline{5-6} 
Eagle                   &                                         &                                             &                                         & \textbf{26.01}         & \textbf{0.58}          \\ \hline
\end{tabular}%
\caption{Impact of different allocation policies on walltime gap reduction and job termination rates for the Fugaku and Eagle workloads.}
\label{tab:allocation_results_combined}
\end{table*}

Direct allocation based on predicted runtimes achieves substantial reductions in the walltime gap, reaching 53--56\% on Fugaku and 55--62\% on Eagle depending on the model. However, this strategy leads to unacceptably high job termination rates, exceeding 41--45\% on Fugaku and 31--44\% on Eagle. Among the evaluated predictors, the NN ensemble consistently achieves the highest gap reduction while maintaining slightly lower job cut rates than the linear and NN-best models.

\begin{figure*}[h]
\centering
\includegraphics[width=0.48\linewidth]{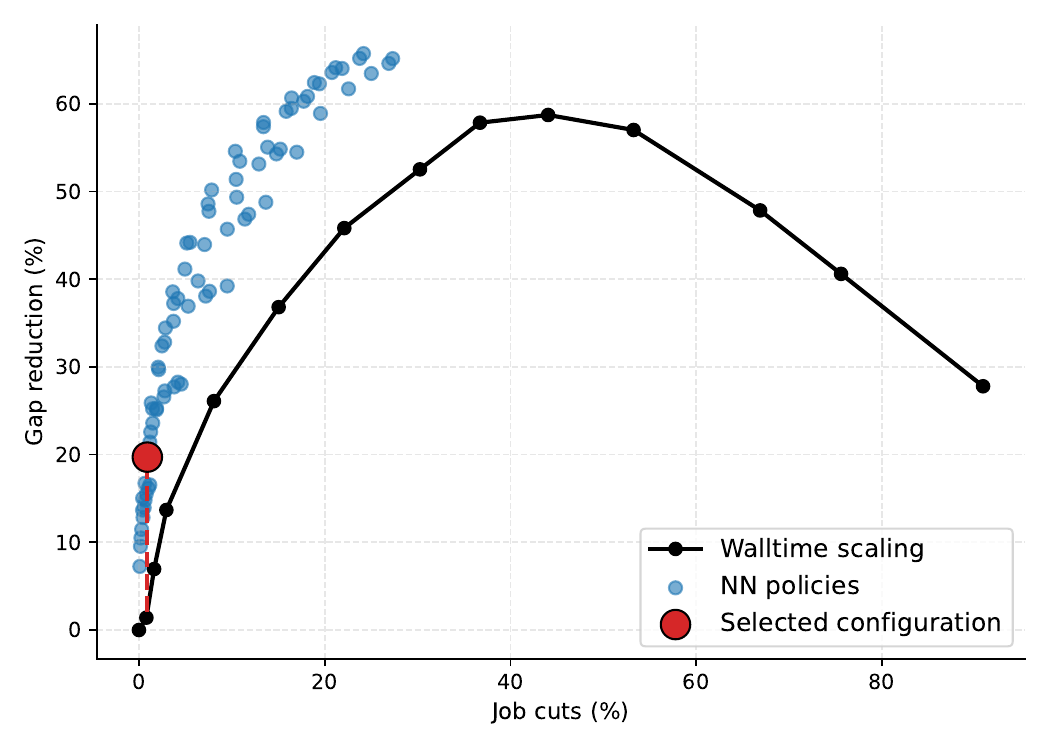}
\includegraphics[width=0.48\linewidth]{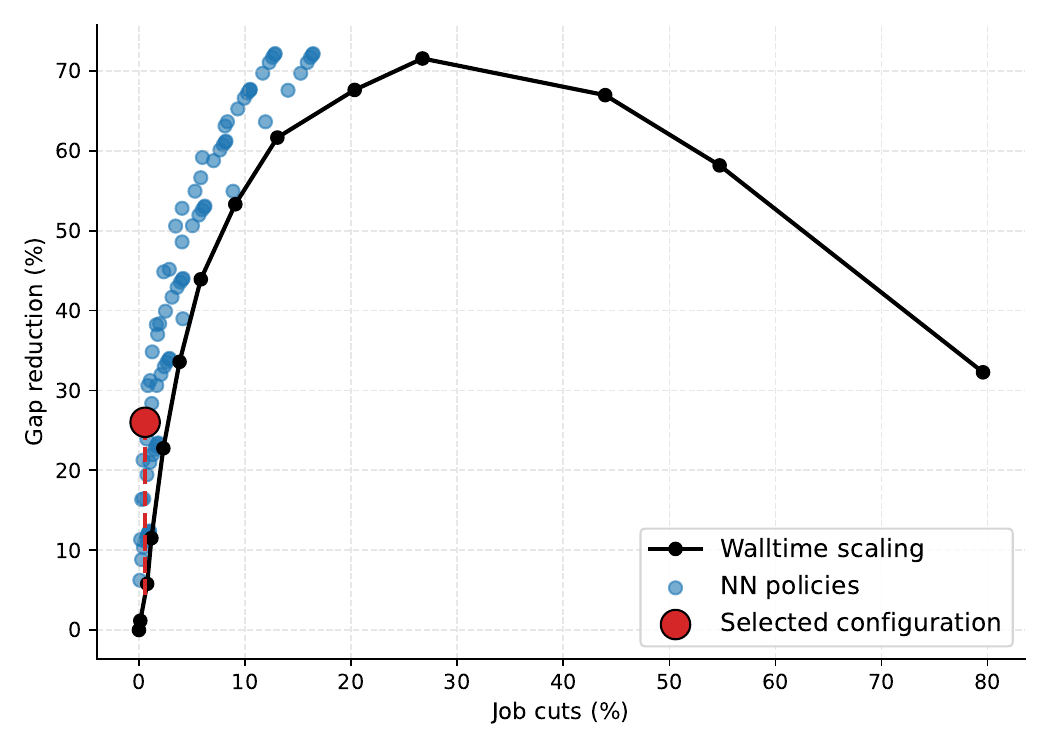}
\caption{Pareto trade-off between walltime gap reduction and job termination rates for the Fugaku (left) and Eagle (right) workloads. 
The black curve corresponds to the walltime scaling baseline evaluated for scaling factors in $[0.01,1.00]$. 
Blue points represent configurations of the proposed similarity-aware SAFE policy using the NN ensemble predictor, with $\alpha \in \{0.1,\dots,0.9\}$ and $\beta \in \{0.1,\dots,0.9\}$. 
The highlighted point indicates the selected configuration ($\alpha=0.1$, $\beta=0.5$).}
\label{fig:pareto_policy}
\end{figure*}

The walltime scaling baseline yields moderate gap reductions while maintaining relatively low job termination rates. Scaling by $0.80 \times elpl$ achieves gap reductions of 26.10\% on Fugaku and 22.75\% on Eagle, while $0.90 \times elpl$ further reduces job cuts at the cost of smaller improvements. As shown in Figure~\ref{fig:pareto_policy}, these configurations correspond to moderate operating points along the baseline trade-off curve.

Applying a multiplicative safety factor of $\kappa=1.5$ to predictions \cite{vallabhajosyula2022towards} achieves the highest gap reductions overall, reaching 63.89\% on Fugaku and 68.30\% on Eagle for the NN  ensemble, while reducing job cuts to around 20\% compared to direct allocation. However, a 20\% job termination rate remains prohibitive for production HPC environments.

The proposed SAFE rule (Contribution~2) significantly reduces job cuts while maintaining moderate gap reduction. Using the NN ensemble, SAFE achieves a gap reduction of 19.19\% on Fugaku and 31.63\% on Eagle, while limiting job terminations to 1.10\% and 1.03\%, respectively.

The similarity-aware SAFE rule (Contribution~3) further reduces job termination rates at a modest cost in gap reduction. On Fugaku, job cuts decrease from 1.10\% to 0.91\%, while the gap reduction slightly increases from 19.19\% to 19.71\%. On Eagle, job cuts decrease from 1.03\% to 0.58\%, accompanied by a reduction in gap improvement from 31.63\% to 26.01\%. Overall, the similarity-aware SAFE policy achieves the lowest job termination rates across both systems, reducing failures below 1\% while still providing substantial walltime gap reduction.

Figure~\ref{fig:pareto_policy} illustrates the trade-off between walltime gap reduction and job termination rates. The black curve represents the baseline strategy that uniformly scales the user-requested walltime, while the blue points correspond to configurations of the proposed similarity-aware SAFE policy using the NN ensemble predictor.

The figure highlights the trade-off between walltime gap reduction and job termination risk. For comparable job cut rates, the similarity-aware SAFE policy consistently achieves larger gap reductions than the walltime scaling baseline, demonstrating that prediction-informed policies can improve walltime efficiency while maintaining low failure risk.

These results demonstrate that the similarity-aware SAFE policy provides the most favorable trade-off between walltime gap reduction and job termination risk across both systems.
\section{Discussion}
\label{sec:discussion}
The results show that directly using runtime predictions for walltime allocation leads to excessive job termination rates. This highlights the practical difficulty of deploying runtime predictors directly in production HPC environments.

The SAFE policies are intentionally designed as conservative bounded adjustment rules that prioritize low job termination rates while still enabling meaningful walltime reduction.

As part of this empirical study, some limitations remain. First, the parameters controlling the SAFE policies were manually selected and were not optimized automatically. Second, the evaluation focuses on walltime estimation rather than full scheduler simulation, meaning that broader system-level effects such as queue waiting times and backfilling efficiency were not explicitly measured. \new{Third, the similarity criterion used to distinguish IN and OUT jobs is based on an exact match of feature signatures; a more graded notion of similarity, for example, a distance threshold in embedding space, could provide finer-grained confidence estimates for borderline cases. Finally, although the framework is validated on two large-scale production HPC systems using different workload managers (PBS and SLURM), evaluating additional systems with different workload characteristics would further assess the generality of the proposed framework.}

Despite these limitations, the results demonstrate that combining runtime prediction with conservative bounded adjustment policies can provide stable trade-offs between walltime reduction and job termination risk across two large-scale production HPC systems.

\new{From an operational perspective, SAFE acts as a deployment layer between runtime prediction and the scheduler. This separation enables improvements in runtime prediction models to be incorporated without modifying the walltime recommendation policy itself. Likewise, scheduler implementations can benefit from improved walltime recommendations without requiring changes to SAFE, making the framework suitable for incremental deployment in production HPC environments.}

\new{Another practical implication of the proposed framework is that it explicitly acknowledges the asymmetric cost of prediction errors in HPC systems. While moderate overestimation mainly reduces scheduling efficiency, underestimation may prematurely terminate jobs and disrupt users' workflows. SAFE therefore deliberately favors conservative bounded adjustments that substantially reduce the operational risk of premature job termination while still improving walltime utilization.}
\section{Conclusions}
\label{sec:conclusion}
\new{This paper presented a predictor-agnostic framework for conservative walltime recommendation in HPC environments, evaluated on large-scale production logs from the Fugaku and Eagle supercomputers comprising more than 35 million jobs.}

We investigated three complementary aspects of safe walltime recommendation. First, a similarity-aware prediction regime that distinguishes repeated workloads (IN) from previously unseen jobs (OUT), revealing consistently higher prediction accuracy for IN jobs across both systems and all evaluated models. Second, a conservative SAFE adjustment rule that translates predicted runtimes into bounded walltime recommendations, limiting job termination rates to approximately 1\% while achieving walltime gap reductions of 19.19\% on Fugaku and 31.63\% on Eagle. Third, a similarity-aware extension of the SAFE rule that applies an additional safety margin to OUT jobs, further reducing job termination rates below 1\% on both systems for the selected configuration, with only a modest impact on gap reduction.

Across both systems, the proposed policies consistently outperform walltime scaling baselines at comparable job termination rates, demonstrating that prediction-informed allocation can improve walltime efficiency without compromising reliability. These results suggest that similarity-aware, bounded adjustment policies are a promising practical approach to walltime recommendation in production HPC environments. 

\new{Because SAFE is predictor-independent, future advances in runtime prediction, including tree ensembles, gradient boosting, uncertainty-aware predictors, or foundation models, can be integrated directly into the framework without modifying the conservative walltime recommendation policy.}

Several limitations remain. The proposed adjustment policies rely on manually selected safety parameters and were not optimized automatically. In addition, the evaluation focuses on walltime estimation rather than full scheduler simulation, meaning that broader system-level effects such as queue waiting times and backfilling efficiency were not explicitly measured.

\new{Future work includes automatic calibration of the SAFE parameters, uncertainty-aware runtime prediction, asymmetric learning objectives that penalize underprediction more heavily, integration of the proposed framework into scheduler simulators to quantify impacts on queue waiting time, backfilling efficiency, throughput, utilization, and job turnaround time, as well as duration-aware prediction approaches, such as specialised models for different runtime regimes or hierarchical predictors.}

\section*{Acknowledgment}
\new{The SEANERGYS project receives funding from the European High-Performance Computing Joint Undertaking (JU) under grant agreement no 101177590.}

\new{The JU receives support from the European Union's Horizon Europe research and innovation programme and Czech Republic, France, Germany, Greece, Italy, and Spain.}

\bibliographystyle{IEEEtran}
\bibliography{biblio}

@inproceedings{zhang2012hpc,
  title={HPC usage behavior analysis and performance estimation with machine learning techniques},
  author={Zhang, Hao and You, Haihang and Hadri, Bilel and Fahey, Mark},
  booktitle={Proceedings of the International Conference on Parallel and Distributed Processing Techniques and Applications (PDPTA)},
  pages={1},
  year={2012},
  organization={The Steering Committee of The World Congress in Computer Science, Computer~…}
}

@article{vercellino2023machine,
  title={A machine learning approach for an HPC use case: The jobs queuing time prediction},
  author={Vercellino, Chiara and Scionti, Alberto and Varavallo, Giuseppe and Viviani, Paolo and Vitali, Giacomo and Terzo, Olivier},
  journal={Future Generation Computer Systems},
  volume={143},
  pages={215--230},
  year={2023},
  publisher={Elsevier}
}

@phdthesis{zrigui2021understanding,
  title={Understanding and improving HPC performance using Machine Learning and Statistical analysis},
  author={Zrigui, Salah},
  year={2021},
  school={Universit{\'e} Grenoble Alpes [2020-....]}
}

@inproceedings{chen2020runtime,
title={Runtime prediction of high-performance computing jobs based on ensemble learning},
author={Chen, Xiaomeng and Zhang, Hui and Bai, Hanli and Yang, Chunming and Zhao, Xujian and Li, Bo},
booktitle={Proceedings of the 2020 4th International Conference on High Performance Compilation, Computing and Communications},
pages={56–62},
year={2020}
}

@inproceedings{tanash2021ampro, title={AMPRO-HPCC: A machine-learning tool for predicting resources on slurm HPC clusters}, author={Tanash, Mohammed and Andresen, Daniel and Hsu, William}, booktitle={ADVCOMP... the... International Conference on Advanced Engineering Computing and Applications in Sciences}, volume={2021}, pages={20}, year={2021} }

@incollection{vallabhajosyula2022towards, title={Towards practical, generalizable machine-learning training pipelines to build regression models for predicting application resource needs on hpc systems}, author={Vallabhajosyula, Manikya Swathi and Ramnath, Rajiv}, booktitle={Practice and Experience in Advanced Research Computing 2022: Revolutionary: Computing, Connections, You}, pages={1--5}, year={2022} }

@inproceedings{gaussier2015improving,
title={Improving backfilling by using machine learning to predict running times},
author={Gaussier, Eric and Glesser, David and Reis, Valentin and Trystram, Denis},
booktitle={Proceedings of the International Conference for High Performance Computing, Networking, Storage and Analysis},
pages={1--10}, year={2015}
}

@article{wang2021rlschert,
title={Rlschert: An hpc job scheduler using deep reinforcement learning and remaining time prediction},
author={Wang, Qiqi and Zhang, Hongjie and Qu, Cheng and Shen, Yu and Liu, Xiaohui and Li, Jing},
journal={Applied Sciences},
volume={11},
number={20},
pages={9448},
year={2021},
publisher={MDPI} }

@article{zaborovsky2023improving,
  title={Improving Efficiency of Hybrid HPC Systems Using a Multi-agent Scheduler and Machine Learning Methods},
  author={Zaborovsky, Vladimir S and Utkin, Lev V and Muliukha, Vladimir A and Lukashin, Alexey A},
  journal={Supercomputing Frontiers and Innovations},
  volume={10},
  number={2},
  pages={104--126},
  year={2023}
}

@incollection{menear2023mastering,
  title={Mastering hpc runtime prediction: From observing patterns to a methodological approach},
  author={Menear, Kevin and Nag, Ambarish and Perr-Sauer, Jordan and Lunacek, Monte and Potter, Kristi and Duplyakin, Dmitry},
  booktitle={Practice and Experience in Advanced Research Computing 2023: Computing for the Common Good},
  pages={75--85},
  year={2023}
}

@inproceedings{gallo2025mitigating,
  title={Mitigating Concept Drift in Job Execution Time Prediction Models},
  author={Gallo, Bernardo and Drummond, L{\'u}cia and Viterbo, Jos{\'e}},
  booktitle={Symposium on Knowledge Discovery, Mining and Learning (KDMiLe)},
  pages={81--88},
  year={2025},
  organization={SBC}
}

@inproceedings{liu2025ora,
title={ORA: Job Runtime Prediction for High-Performance Computing Platforms Using the Online Retrieval-Augmented Language Model},
author={Liu, Hongyi and Ma, Yinping and Huang, Xiaosong and Zhang, Lingzhe and Jia, Tong and Li, Ying},
booktitle={Proceedings of the 39th ACM International Conference on Supercomputing},
pages={884--894},
year={2025}
}

@misc{OEDI_Dataset_5860,
title = {NREL Eagle supercomputer jobs},author = {Duplyakin, Dmitry and Menear, Kevin},abstractNote = {High performance computing dataset with 11M+ jobs from NREL's Eagle supercomputer. These jobs were submitted to run on Eagle between Nov 2018 and Feb 2023. The data are sufficiently anonymized and do not include sensitive user or project data. HPC research community does not have many public, large, and complete job traces like this one, and releasing this dataset should help address this gap.},url = {https://data.openei.org/submissions/5860},year = {2023},howpublished = {Open Energy Data Initiative (OEDI), National Renewable Energy Laboratory, https://data.openei.org/submissions/5860},note = {Accessed: 2026-02-18}
}

@misc{antici_2024_11467483,
  author = {Antici, Francesco and others},
  title = {F-DATA: A Fugaku Workload Dataset for Job-Centric Predictive Modelling in HPC Systems},
  year = {2024},
  doi = {10.5281/zenodo.11467483},
  url = {https://doi.org/10.5281/zenodo.11467483}
}

@article{li2021okcm,
  title={OKCM: improving parallel task scheduling in high-performance computing systems using online learning: J. Li et al.},
  author={Li, Jingbo and Zhang, Xingjun and Han, Li and Ji, Zeyu and Dong, Xiaoshe and Hu, Chenglong},
  journal={The Journal of Supercomputing},
  volume={77},
  number={6},
  pages={5960--5983},
  year={2021},
  publisher={Springer},
  doi={10.1007/s11227-020-03506-5}
}

@article{FEITELSON20142967,
title = {Experience with using the Parallel Workloads Archive},
journal = {Journal of Parallel and Distributed Computing},
volume = {74},
number = {10},
pages = {2967-2982},
year = {2014},
issn = {0743-7315},
doi = {https://doi.org/10.1016/j.jpdc.2014.06.013},
url = {https://www.sciencedirect.com/science/article/pii/S0743731514001154},
author = {Dror G. Feitelson and Dan Tsafrir and David Krakov},
}

@article{antici2025f,
  title={F-data: A Fugaku workload dataset for job-centric predictive modelling in HPC systems},
  author={Antici, Francesco and Bartolini, Andrea and Domke, Jens and Kiziltan, Zeynep and Yamamoto, Keiji},
  journal={Scientific Data},
  volume={12},
  number={1},
  pages={1321},
  year={2025},
  publisher={Nature Publishing Group UK London}
}

@article{choi2026uarp,
  title={UARP: uncertainty-aware runtime prediction for preventing scheduler termination under Wallclock constraints in HPC},
  author={Choi, Jiheon and Oh, Sangyoon},
  journal={The Journal of Supercomputing},
  volume={82},
  number={5},
  pages={292},
  year={2026},
  publisher={Springer}
}

@inproceedings{cui2024clustering,
  title={Clustering based job runtime prediction for backfilling using classification},
  author={Cui, Hang and Takahashi, Keichi and Shimomura, Yoichi and Takizawa, Hiroyuki},
  booktitle={Workshop on Job Scheduling Strategies for Parallel Processing},
  pages={40--59},
  year={2024},
  organization={Springer}
}

@inproceedings{hai2026fine,
  title={Fine-tuned sentence-bert for hpc job outcome prediction via textual feature embedding},
  author={Hai, Thanh Hoang Le and Tuan, Huy Nguyen and Dang, Bao Tran and Thuong, Bao Vo and Thoai, Nam},
  booktitle={Proceedings of the Supercomputing Asia and International Conference on High Performance Computing in Asia Pacific Region Workshops},
  pages={378--387},
  year={2026}
}

\end{document}